\documentclass[aps,prb,groupedaddress,showpacs,notitlepage,twocolumn]{revtex4-1}
\usepackage{hyperref}
\usepackage{amsmath}
\usepackage{graphicx}
\usepackage{color}

\newcommand{\be}{\begin{equation}}
\newcommand{\ee}{\end{equation}}

\begin{document}
\title{Flat bands in lattices with non-Hermitian coupling}

\author{Daniel Leykam$^{1}$, Sergej Flach$^{2}$, and Y. D. Chong$^{1,3}$}
%\email[Email: ]{dleykam@ntu.edu.sg}

\affiliation{$^{1}$Division of Physics and Applied Physics, School of Physical and Mathematical Sciences,\\
Nanyang Technological University, Singapore 637371, Singapore \\
$^{2}$Center for Theoretical Physics of Complex Systems, Institute for Basic Science, Daejeon, South Korea \\
$^{3}$Centre for Disruptive Photonic Technologies, Nanyang Technological University, Singapore 637371, Singapore}
\date{\today}

\begin{abstract}
We study non-Hermitian photonic lattices that exhibit competition between conservative and non-Hermitian (gain/loss) couplings. A bipartite sublattice symmetry enforces the existence of non-Hermitian flat bands, which are typically embedded in an auxiliary dispersive band and give rise to non-diffracting ``compact localized states''.  Band crossings take the form of non-Hermitian degeneracies known as exceptional points.  Excitations of the lattice can produce either diffracting or amplifying behaviors. If the non-Hermitian coupling is fine-tuned to generate an effective $\pi$ flux, the lattice spectrum becomes completely flat, a non-Hermitian analogue of Aharonov-Bohm caging in which the magnetic field is replaced by balanced gain and loss. When the effective flux is zero, the non-Hermitian band crossing points give rise to asymmetric diffraction and anomalous linear amplification.
\end{abstract}

%\pacs{63.20.Pw, 42.82.Et, 78.67.Pt}
%63.20.Pw; localized modes
%42.82.Et; waveguides, couplers, and arrays
%78.67.Pt; multilayers, superlattices, photonic structures

\maketitle

\section{Introduction}

Flat bands are one of a handful of band structure features, alongside band gaps and band-crossing points, that are of key interest in photonic, electronic, and cold atom lattices \cite{bergholtz_review,derzhko_review}.  In a flat band, diffraction is suppressed due to destructive interference, in a manner analogous to geometric frustration \cite{frustration}, giving rise to eigenmodes that are compactly localized in space.  Flat bands have been observed with cold atom lattices~\cite{optical_lattice}, coupled laser arrays~\cite{coupled_lasers}, superconducting wire networks~\cite{ab_cage,ab_cage_experiment}, and photonic lattices~\cite{vicencio2015,mukherjee2015,diamond_ladder,polariton_flat_band,sawtooth}.  Lattices containing flat bands provide a unique setting for exploring unconventional localization, anomalous magnetic phases, and strongly-correlated states of matter such as fractional quantum Hall effects and spin liquids~\cite{frustration,derzhko_review,bergholtz_review}.  

Photonic lattices have the distinction of easily supporting structured gain (amplification) and loss (dissipation), which can be used to study interesting non-Hermitian wave effects~\cite{pt_lattice_1,pt_lattice_3,pt_lattice_4,pt_optical_lattice,zhen2015,cerjan2016}.  For instance, non-Hermitian lattices typically have non-orthogonal eigenstates and complex band energies; but when gain and loss are distributed so that the lattice is invariant under a combination of parity and time-reversal ($\mathcal{PT}$ symmetry), it is possible to have bands that are purely real~\cite{bender,PT_review,PT_review_2}.  Moreover, $\mathcal{PT}$ symmetric band structures can exhibit non-Hermitian degeneracies known as exceptional points (EPs), where a pair of eigenstates coalesce and the Hamiltonian becomes defective~\cite{pt_breaking, pt_optics, EP, exceptional_point, ding_2015, EP_acoustic}.  EPs can give rise to a wide variety of interesting phenomena, including unidirectional invisibility, chiral lasing, and enhanced spontaneous emission~\cite{invisibility,invisibility_2,EP_laser,EP_laser_2,lin2016}.

An important question is whether the existence of flat bands is compatible with non-Hermiticity, and if so whether the presence of EPs might alter the behavior of the flat band states. Previous studies have analyzed how Hermitian flat bands are changed by the application of non-Hermitian perturbations \cite{pt_flat,chern2015,molina2015}, finding either that the symmetries protecting the flat band states are spoilt, or that the flat bands simply acquire nonzero gain or loss (breaking $\mathcal{PT}$ symmetry). Recently, Ramezani has shown that a flat band with completely real eigenvalues can exist in a quasi-1D $\mathcal{PT}$ photonic lattice \cite{PT_FB}.  This $\mathcal{PT}$ symmetric flat band was realized by fine-tuning gain/loss levels, and it is not obvious what features of this model, compared to earlier models \cite{pt_flat,chern2015,molina2015}, make it possible.

Here, we show that $\mathcal{PT}$ symmetric flat bands generically occur in non-Hermitian lattices with a bipartite sublattice symmetry hosting a differing number of sites per sublattice~\cite{lieb,chiral_FB}.  The flat bands emerge from competition between incompatible Hermitian and non-Hermitian (gain/loss) couplings, and possess properties that are qualitatively different from their Hermitian counterparts.  We identify two novel types of $\mathcal{PT}$ symmetric flat band:

(i) A non-Hermitian variant of an ``Aharonov-Bohm cage'' (wave localization induced by a fine-tuned magnetic flux)~\cite{ab_cage,ab_cage_experiment,ab_cage_longhi}, with gain/loss couplings playing the role of a complex magnetic vector potential.  The Bloch Hamiltonian is defective for \emph{every} wavevector, and transport is entirely suppressed.  Certain states experience anomalous sub-exponential (quadratic) amplification.

(ii) A flat band embedded within another dispersive band, where the band crossing points form isolated EPs embedded in a continuum.  Unlike previously-studied ``embedded EPs'' created by localized defects~\cite{EP_continuum}, the EPs in this case occur in a translationally invariant lattice, corresponding to unstable Bloch waves displaying sub-exponential amplification. 

In both cases, we find numerically that the flat band-induced wave localization is robust to weak disorder.  The flat band states remain strongly localized, but the eigenvalue spectrum becomes complex because the disorder breaks the $\mathcal{PT}$ symmetry.  As a result, some of the flat band states become preferentially amplified.

We will consider a photonic lattice of coupled optical waveguides.  Under the tight-binding approximation, the evolution of the optical field within the lattice can be described by $i \partial_z |\Psi \rangle = \hat{H} |\Psi \rangle$, where $z$ is the position along the waveguide axis and $\hat{H}$ is a Hamiltonian matrix.  The diagonal terms of $\hat{H}$ describe the propagation constant and gain/loss of the $n$th site (waveguide), and the off-diagonal terms $\kappa_{nm}$ describe the coupling between the $n$th and $m$th sites.

Non-Hermitian couplings, $\kappa_{nm} \ne \kappa_{mn}^*$, describe a situation where the mode amplitude undergoes gain or loss while hopping between sites.  One way to realize such couplings physically is to embed amplifying or lossy media between adjacent waveguides~\cite{active_coupler,dark_state_laser,longhi_gauge,dark_laser_experiment}.  When neighboring waveguides are in phase, the evanescent tails of the waveguide modes interfere constructively, enhancing the intensity within the gain (lossy) medium and leading to stronger amplification (attenuation). Conversely, when the neighbouring waveguides are out of phase, weaker amplification (attenuation) occurs. Thus, the effective gain (loss) is sensitive to the relative phase between neighboring waveguides.  Non-Hermitian couplings can also be realized by periodically modulating the on-site gain/loss~\cite{invisibility}, or as an effective description of larger waveguide networks~\cite{longhi2016,diffusive}.

In this paper, we will assume for simplicity that the couplings are symmetric, $\kappa_{nm} = \kappa_{mn}$.  Each coupling term consists of a Hermitian part, $\mathrm{Re}(\kappa_{nm})$, and a non-Hermitian part, $\mathrm{Im}(\kappa_{nm})$.  It is the competition or ``frustration'' between the Hermitian and non-Hermitian parts that will produce the $\mathcal{PT}$ symmetric flat bands.

\section{Frustrated trimer}
\label{sec:trimer}

\begin{figure}

\includegraphics[width=\columnwidth]{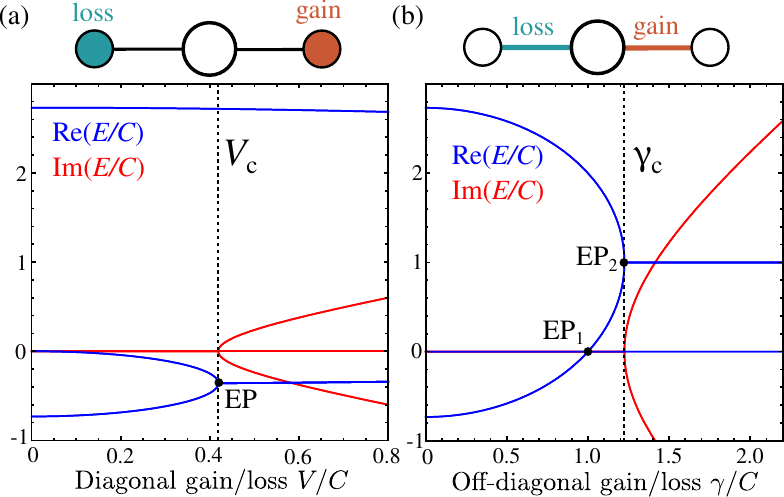}

\caption{Schematic of trimers with diagonal and off-diagonal gain/loss and their energy spectra. (a) Diagonal gain/loss $V$: red (blue) sites have net gain (loss). Complex energy eigenvalues $E$ emerge at $V_c \approx 0.42C$ when neighboring modes coalesce at EP. (b) Non-Hermitian coupling $\gamma$: red (blue) links indicate net gain (loss) occurs when neighboring sites are in phase, and vice versa when out of phase. Complex energy eigenvalues emerge at $\gamma_c \approx 1.22C$ via EP$_2$ only after first crossing EP$_1$. Here $\Delta = 2C$.}

\label{fig:frustration}

\end{figure}

The simplest examples of geometrical frustration involve three modes, e.g.~trimers of antiferromagnetic spins~\cite{frustration}.  Similarly, we begin our study of competing non-Hermitian couplings by analyzing a $\mathcal{PT}$-symmetric trimer, described by the Hamiltonian matrix
\be 
\hat{H} = \left( \begin{array}{ccc} i V & C + i \gamma & 0 \\ C + i \gamma & \Delta & C - i \gamma \\ 0 & C - i \gamma & - i V \end{array} \right). \label{eq:full_trimer}
\ee
When the central site detuning $\Delta$ is real, this is the most general Hamiltonian describing a $\mathcal{PT}$ symmetric trimer that is reflection symmetric ($\mathcal{P}$) about the central site and has only nearest-neighbor coupling. The real parameter $V$ describes the on-site gain and loss, while the non-Hermitian couplings are $\kappa_{12} = \kappa_{21} = C + i \gamma = \kappa_{23}^* = \kappa_{32}^*$.

As is usual for $\mathcal{PT}$-symmetric Hamiltonians, $\hat{H}$ possesses a ``$\mathcal{PT}$-unbroken phase'': when the non-Hermitian parameters $V$ and $\gamma$ are small, all the eigenvalues are real.  Complex eigenvalues emerge abruptly as we tune the system through an EP by increasing the magnitude of $V$ and/or $\gamma$. The eigenstates of a $\mathcal{PT}$-symmetric system form complex conjugate pairs, and the $\mathcal{PT}$-breaking transition involves \emph{pairs} of eigenstates~\cite{bender,PT_review,PT_review_2}. In a system with an odd number of eigenstates, there are two distinct ways for $\mathcal{PT}$-breaking to occur, which we illustrate in Fig.~\ref{fig:frustration} using the limiting cases $\gamma = 0$ and $V=0$.  In the first case [Fig.~\ref{fig:frustration}(a)], where the gain/loss is purely on-site, a pair of neighboring eigenmodes coalesce to produce a ``conventional'' $\mathcal{PT}$-breaking transition at an EP, at $V = V_c \approx 0.42C$.  The remaining mode always has a real eigenvalue, and does not play any important role in the transition. In this case, the $\mathcal{PT}$-breaking can be captured entirely by a two-mode approximation (see below).

A different kind of $\mathcal{PT}$-breaking transition is shown in Fig.~\ref{fig:frustration}(b).  Here, the gain/loss is purely off-diagonal, and the system enters the $\mathcal{PT}$-broken phase at EP$_2$ when $\gamma = \gamma_c = \sqrt{C^2 + \Delta^2/8}$, where the first and third eigenmodes coalesce. Before reaching EP$_2$, however, the second and third eigenmodes must first cross at the point labelled EP$_1$ ($\gamma = C$).  This degeneracy point is an EP, because the two participating eigenmodes coalesce to $| \Psi_2 \rangle = | \Psi_3 \rangle = (e^{-i \pi/4}, 0, -e^{i \pi/4})$, and $\hat{H}$ is defective. However, EP$_1$ does not correspond to a $\mathcal{PT}$-breaking transition; all the eigenvalues in its vicinity are real.

To gain a better understanding of EP$_1$, note that $\hat{H}$ has a ``dark eigenstate'' $|\Psi_2 \rangle = (C-i\gamma,0,-C-i\gamma)$ in which the central site is unexcited. Its eigenvalue $E_2 = 0$ is independent of $C,\gamma$, and $\Delta$; hence it cannot belong to a pair of $\mathcal{PT}$-broken eigenvalues. In the Hermitian context, this kind of ``flat'' eigenvalue is characteristic of systems with a bipartite symmetry, where pairs of identical states are only coupled via an intermediate state~\cite{lieb,chiral_FB}.  In the present system, $\gamma/C$ fixes the relative phase of the two end-sites, which is reminiscent of the role of a vector potential. The relative phase is required to maintain a balanced flow of energy between the gain and lossy media. But in contrast to Hermitian vector potentials in 1D arrays, this relative phase cannot simply be removed by a gauge transformation; it has the observable effect of inducing eigenstate non-orthogonality. The dark state is $\mathcal{P}$-antisymmetric in the Hermitian limit ($\gamma = 0$) and $\mathcal{P}$-symmetric in the limit of completely non-Hermitian couplings ($C=0$).

In the strongly-detuned limit, $\Delta \gg C,V,\gamma$, the central site may be eliminated from the eigenvalue problem. For small energies $\delta E \ll \Delta$, Eq.~(\ref{eq:full_trimer}) reduces to an effective $2\times2$ Hamiltonian:
\be 
\hat{H}_{\mathrm{eff}} = \left( \begin{array}{cc} -(C+i\gamma)^2/\Delta +iV & -(C^2+\gamma^2)/\Delta \\ -(C^2+\gamma^2)/\Delta & -(C-i\gamma)^2/\Delta -iV \end{array} \right). \label{eq:effective_ham}
\ee
For $\gamma = 0$, $\hat{H}_{\mathrm{eff}}$ is the Hamiltonian that describes a $\mathcal{PT}$-symmetric dimer with reduced effective coupling strength $|C_{\mathrm{eff}}| = C^2/\Delta$.  This has a $\mathcal{PT}$-breaking transition at $|V| = |C_{\mathrm{eff}}|$, corresponding to the point labelled ``EP'' in Fig.~\ref{fig:frustration}(a).  On the other hand, the ``frustrated'' trimer ($V=0$) has an effective coupling strength $|C_{\mathrm{eff}}| = (C^2+\gamma^2)/\Delta$ that increases with the non-Hermiticity $\gamma$, in competition with the effective non-Hermitian potential $V_{\mathrm{eff}} = -C\gamma/\Delta$. This induces an EP at $|\gamma| = |C|$, corresponding to EP$_1$ in Fig.~\ref{fig:frustration}(b), but the perturbative eigenvalues $\delta E = 0, 2(\gamma^2-C^2)/\Delta$ remain real; $\mathcal{PT}$-breaking cannot occur within the two mode approximation.

This simple model demonstrates that a bipartite $\mathcal{PT}$-symmetric system, with competing Hermitian and non-Hermitian couplings, produces a dark state with energy pinned to $E_0 = 0$. Even when the bipartite symmetry is broken (i.e., $V$ and $\gamma$ both nonzero), the $\mathcal{PT}$-breaking is limited to $C-\sqrt{|\Delta V|} < |\gamma| < C+\sqrt{|\Delta V|}$.  For sufficiently large $\gamma$, the system returns to the unbroken phase. 

\section{Frustrated lattices}
\label{sec:lattice}

We now generalize the trimer to a $\mathcal{PT}$-symmetric lattice with bipartite sublattice symmetry, where the dark state becomes a flat band of zero-energy Bloch states. Consider a 1D lattice obeying tight-binding equations
\begin{subequations}
\label{eq:rspace}
\begin{align}
E a_n &= \kappa_1 b_n + \kappa_2 b_{n+1}, \label{eq:rspace1} \\
E b_n &= \Delta b_n + \kappa_1 a_n + \kappa_2 a_{n-1} + \kappa_3 c_n + \kappa_4 c_{n-1},
\label{eq:rspace2} \\
E c_n &= \kappa_3 b_n + \kappa_4 b_{n+1}, \label{eq:rspace3}
\end{align}
\end{subequations}
where $n=1,...,N$; $\Delta$ is the detuning of the central sites; and $\kappa_j \equiv C_j + i \gamma_j$ are complex couplings. Fourier transforming gives $\hat{H}(k) \Psi(k) = E(k) \Psi(k)$, where
\be 
\hat{H}(k) = \left( \begin{array}{ccc} 0 & \kappa_1  + \kappa_2 e^{i k} & 0 \\ \kappa_1 + \kappa_2 e^{-ik} & \Delta & \kappa_3 + \kappa_4 e^{-ik} \\ 0 & \kappa_3  + \kappa_4 e^{i k} & 0 \end{array} \right). \label{hamiltonian}
\ee
The energy eigenvalues are
\be 
E(k) = 0, \frac{\Delta}{2} \pm \sqrt{ \frac{\Delta^2}{4} + 2 (\kappa_1 \kappa_2 + \kappa_3 \kappa_4) \cos k + \sum_j \kappa_j^2}, \label{energy_spectrum}
\ee
where $C_{\mathrm{eff}} = \kappa_1 \kappa_2 + \kappa_3 \kappa_4$ is an effective coupling strength for two dispersive bands, and $\sum_j \kappa_j^2$ is an effective energy detuning.  The $E=0$ flat band is a generalization of the dark state of the trimer model from Section~\ref{sec:trimer}.  The Bloch states in this band all have $b_n=0$, regardless of the values of $\Delta$, $\kappa_j$, and $k$.  This is due to the bipartite sublattice symmetry in Eqs.~\eqref{eq:rspace1} and \eqref{eq:rspace3}; since Eq.~\eqref{eq:rspace2} only provides $N$ constraints to the $2N$ remaining degrees of freedom ($a_n,c_n$), there are $N$ degenerate solutions.

The flat band allows for the construction of compact localized eigenstates, which are non-diffracting and vanish identically outside of a few sites~\cite{leykam_FB}.  To find these solutions, we assume that the $a_n$ and $c_n$ amplitudes are only nonzero at sites $p,p+1$, and solve Eq.~\eqref{eq:rspace2}:
\begin{align*}
  \kappa_1 a_p + \kappa_3 c_p &= 0\\
  \kappa_2 a_{p+1} + \kappa_4 c_{p+1} &=0 \\
  \kappa_2 a_p + \kappa_4 c_p + \kappa_1 a_{p+1} + \kappa_3 c_{p+1} &= 0.
\end{align*}
This yields the eigenmode amplitudes
\begin{align}
  \begin{aligned}
    a_p &= 1, \qquad\quad\;\; a_{p+1} = \kappa_4/\kappa_3, \\
    c_p &= -\kappa_1/\kappa_3, \;\;\; c_{p+1} = -\kappa_2/\kappa_3, \label{eq:CLS}
  \end{aligned}
\end{align}
with all other $a_n,c_n$ vanishing. 

Similar to the dark mode of the trimer model, the flat band exists independent of whether the $\kappa$ parameters are real (Hermitian) or complex (non-Hermitian).  In the latter case, however, the flat band states have a nontrivial phase profile, and there can be two qualitatively different types of crossings between the flat and dispersive bands, depending on the lattice geometry.

\begin{figure}

\includegraphics[width=\columnwidth]{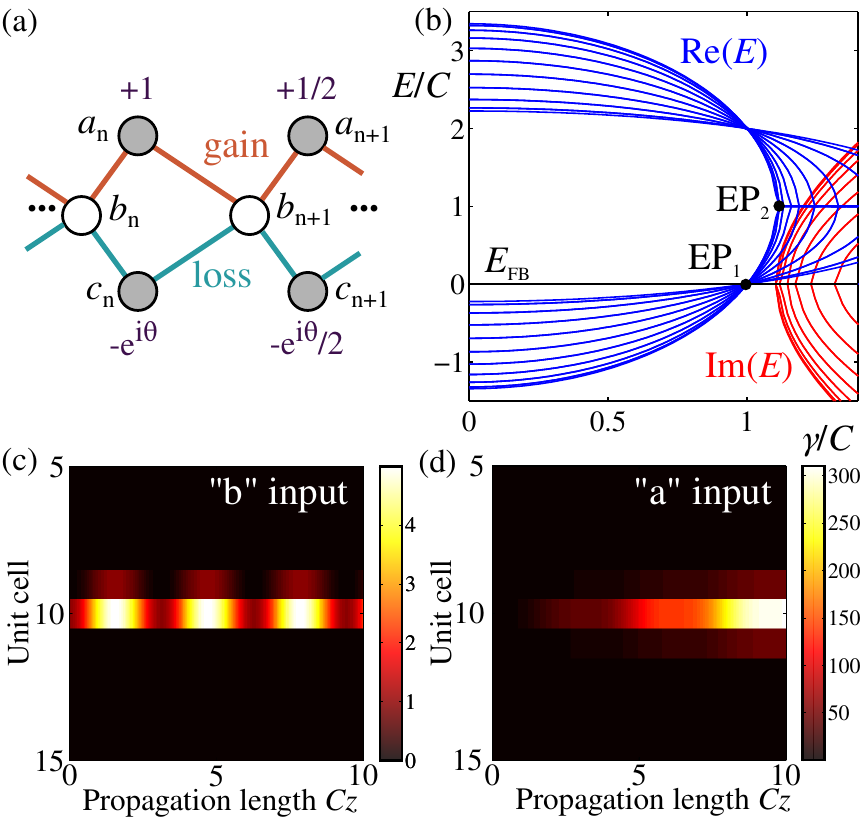}

\caption{(a) Diamond chain of resonators with antisymmetric non-Hermitian couplings, which form a non-Hermitian Aharonov-Bohm cage. Compact localized eigenmode amplitudes with phases $\theta = 2\mathrm{Arg}(C+i\gamma)$ are indicated by shaded sites. (b) Plot of band energies, given by Eq.~\eqref{energy_spectrum}, versus the non-Hermiticity parameter $\gamma/C$.  The spectrum flattens with increasing $\gamma/C$; complex eigenvalues with nonzero $\mathrm{Im}(E)$ (red curves) appear only after the two dispersive bands merge at EP$_2$, $\gamma/C \approx 1.1$.  The other model parameters are $\alpha=1/2$, $\Delta=2C$, and lattice size $N=20$ cells, with periodic boundary conditions. (c,d) Intensity profile $|\Psi_n(z)|^2$ of a localized $b$ or $a$ sublattice excitation, when the lattice is tuned to EP$_1$.  (c) The $b$ input excites an eigenstate that oscillates periodically. (d) The $a$ input leads to quadratically-growing total intensity.}

\label{fig:lattice}

\end{figure}

We first consider the lattice in Fig.~\ref{fig:lattice}(a), which contains two legs with opposite non-Hermitian coupling strengths: $\kappa_1 = C + i \gamma$, $\kappa_2 = \alpha \kappa_1$, $\kappa_3 = \kappa_1^*$, and $\kappa_4 = \kappa_2^*$ (where $\alpha$ is a real parameter).  Between adjacent central ($b_n$) sites, coupling via the upper leg accumulates a phase $\theta = \mathrm{Arg}(\kappa_1 \kappa_2)=2\tan (\gamma / C)$, while coupling via the lower leg accumulates an equal and opposite phase $-\theta$. This resembles a net magnetic flux of $\Delta \theta = 2\theta$.  As we increase $\gamma$ from zero to $C$, the effective coupling $C_{\mathrm{eff}} = \alpha (C^2-\gamma^2$) vanishes, and the dispersive bands consequently all flatten out.  This corresponds to the case $\Delta \theta = \pi$, and is thus analogous to the formation of an ``Aharonov-Bohm cage'' in a Hermitian lattice~\cite{ab_cage,ab_cage_experiment,ab_cage_longhi}. But unlike those Hermitian models, one of the dispersive bands coalesces onto the $E=0$ flat band, forming an entire band of $N$ non-Hermitian degeneracies, as shown in Fig.~\ref{fig:lattice}(b).  At the point EP$_1$ ($\gamma = C$), $\hat{H}(k)$ is not diagonalizable for \emph{any} $k$: it only has two eigenvectors,
\be 
|\Psi_1 \rangle = \left(i, \frac{\Delta}{C}\frac{e^{i\pi/4}}{2+e^{ik}},1\right), \quad  |\Psi_2 \rangle = (i,0,1),
\ee
where $|\Psi_2 \rangle$ is doubly degenerate. Upon increasing $\gamma$ past EP$_1$, the degeneracy is lifted, but the spectrum remains purely real until the two dispersive bands coalesce at EP$_2$ ($\gamma/C \approx 1.1$). 

An interesting property of Aharonov-Bohm cages is that despite the nonzero couplings between lattice sites, localized excitations only spread a finite distance; propagation over longer distances is completely suppressed (not just exponentially suppressed) by destructive interference between the two legs formed by the $a$ and $c$ sites. To test whether our non-Hermitian model has this property, we simulate the evolution $i \partial_z |\Psi\rangle = \hat{H}|\Psi\rangle$ for an initial excitation localized to either an $a$ or $b$ site of the lattice.  The results, shown in Fig.~\ref{fig:lattice}(c,d), show a complete suppression of diffraction when $\gamma = C$.  The initial excitation spreads to the neighboring unit cells, but no further; the intensity beyond the neighbouring cells vanishes to within numerical precision. When eigenstates of $\hat{H}$ are excited, there are periodic oscillations, but otherwise the EP results in quadratic amplification of the total power~\cite{EP_continuum}. By contrast, if we excite four sites according to Eq.~\eqref{eq:CLS}, we can form a compact localized state with $z$-invariant intensity.

\begin{figure}

\includegraphics[width=\columnwidth]{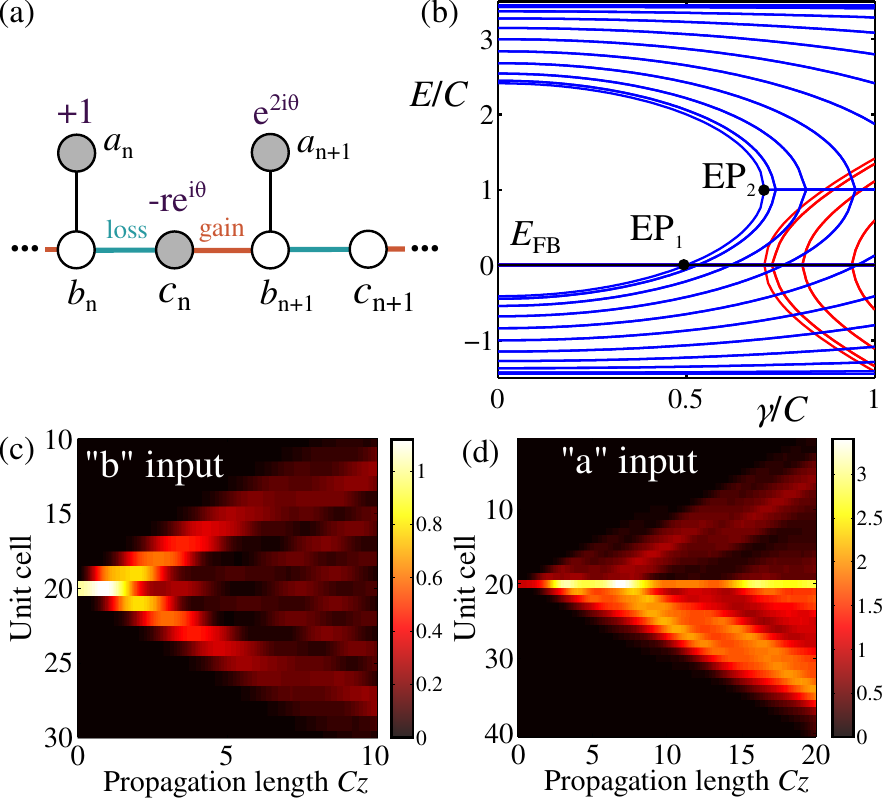}

\caption{(a) Stub lattice hosting a non-Hermitian flat band embedded in a dispersive band. Compact localized eigenmode amplitudes with phase $\theta = \mathrm{Arg}(C+i\gamma)$ are indicated by shaded sites (here, $r^2 \equiv [1+\gamma^2/C^2]^{-1}$). (b) Plot of band energies, given by Eq.~\eqref{energy_spectrum}, versus the non-Hermiticity parameter $\gamma/C$.  Complex eigenvalues (nonzero $\mathrm{Im}(E)$ in red) emerge only when the two dispersive bands merge at EP$_2$, first for modes at the Brillouin zone edge ($k = \pm \pi$) and progressing towards $k=0$.  We take $\Delta=2C$, and a lattice size of $N=20$ cells with periodic boundary conditions. (c,d) Intensity profile $|\Psi_n(z)|^2$ of a localized $b$ or $a$ sublattice excitation, when the lattice is tuned to $\gamma=2C/3$. (c) The $b$ excitation excites dispersive bands, producing discrete diffraction. (d) The $a$ excitation excites both flat and dispersive bands, producing asymmetric diffraction and linear intensity amplification.}

\label{fig:embedded_FB}

\end{figure}

Qualitatively different behavior occurs when each leg has balanced non-Hermitian couplings, such that the effective flux always vanishes. As an example, we consider the ``stub'' lattice~\cite{polariton_flat_band} shown in in Fig.~\ref{fig:embedded_FB}, where coupling occurs via a single leg with balanced gain and loss: $\kappa_1=C$, $\kappa_2=0$, and $\kappa_3=\kappa_4^* = C+i\gamma$.  In this case, the non-Hermiticity parameter $\gamma$ has the effect of increasing the effective coupling $C_{\mathrm{eff}} = C^2 + \gamma^2$, broadening the two dispersive bands. This can be understood as different wavevectors $k$ experiencing different effective amounts of gain/loss. By increasing $\gamma$, we can continuously tune the lower dispersive band eigenvalues in Fig.~\ref{fig:embedded_FB} through the $E=0$ flat band to generate a succession of EP$_1$'s at critical momenta $k = k_c$, where
\be 
\cos (k_c) = 1 - 5/(2+2\gamma^2/C^2).
\ee
The EP$_1$s form isolated exceptional points in the continuum: while the Hamiltonian becomes defective at $k_c$, the spectrum in the vicinity of the EP remains purely real. Previously, similar EPs in the continuum were obtained using a specially-tailored defect potential~\cite{EP_continuum}.  These EPs, however, occur in a periodic lattice.

In Figs.~\ref{fig:embedded_FB}(c,d), we simulate propagation in the lattice below its EP$_2$ threshold, where the isolated EP$_1$s coexist with an entirely real spectrum.  As shown in Fig.~\ref{fig:embedded_FB}(c), a $b$ sublattice input does not excite the flat band or the EP$_1$s.  It instead generates a conventional discrete diffraction pattern, with two ballistically expanding lobes and bounded total power, as required for a $\mathcal{PT}$-symmetric lattice in the unbroken phase.  On the other hand, in Fig.~\ref{fig:embedded_FB}(d) we see that a localized $a$ sublattice input excites a superposition of flat and dispersive band states, including the EP$_1$ states at $\pm k_c$.  The resulting diffraction pattern is strongly asymmetric, due to the amplification at the EP, and there is a residual localized flat band component. Interestingly, the observed power growth is only linear in $z$, slower than the quadratic growth observed in Fig.~\ref{fig:lattice}(d), as well as in previous examples of EPs in the continuum formed by localized defects~\cite{EP_continuum}.  (It is also slower than the exponential growth generated by the complex eigenvalues of unfrustrated $\mathcal{PT}$-symmetric lattices~\cite{pt_lattice_1,pt_lattice_3,pt_lattice_4}.)  Here, the instability is limited to the wavevectors $\pm k_c$, i.e., quadratic amplification is a \emph{collective} effect requiring excitation of all $N$ cells. 

\section{Disorder} 

In practical implementations, fabrication disorder may broaden the flat band to some nonzero width, lifting its degeneracy. We now study how the flat band states behave in this situation. With a random onsite potential, the tight binding equations~\eqref{eq:rspace} become
\begin{align}
E a_n &= \epsilon_n^a a_n + \kappa_1 b_n + \kappa_2 b_{n+1}, \nonumber  \\
E b_n &= (\epsilon_n^b + \Delta) b_n + \kappa_1 a_n + \kappa_2 a_{n-1} + \kappa_3 c_n + \kappa_4 c_{n-1}, \nonumber \\
E c_n &= \epsilon_n^c c_n + \kappa_3 b_n + \kappa_4 b_{n+1}.\label{eq:rspace_disorder}
\end{align}
%\end{subequations}
The uniformly-distributed independent random variables $\epsilon_n^j \in [-W/2,W/2]$ describe uncorrelated on-site disorder in the waveguide depths, or in the resonant frequencies of ring resonators~\cite{longhi_gauge}. The disorder breaks the sublattice and $\mathcal{PT}$ symmetries of the non-Hermitian flat band lattice, so the energy eigenvalues $E$ all become non-real.

\begin{figure}

\includegraphics[width=\columnwidth]{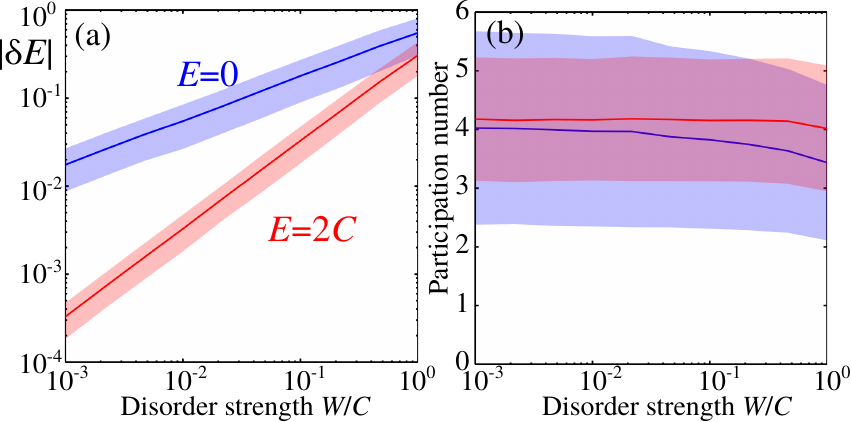}

\caption{Disordered non-Hermitian Aharonov-Bohm cage. (a) Mean eigenvalue shifts $\langle |\delta E | \rangle$ versus disorder strength $W$, averaging over 200 disorder samples.  The non-Hermitian ($E=0$; blue) and Hermitian-like ($E=2C$; red) flat bands scale as $\sim \sqrt{W}$ and $\sim W$ respectively. Shaded regions denote the standard deviation. (b) Mean participation number. All the eigenmodes are strongly localized, $P \sim P_{\mathrm{CLS}}$, and the localization is insensitive to $W$. Here $\gamma = C$ and $\Delta = 2C$, as in Fig.~\ref{fig:lattice}, and the lattice size is $N=40$ units cells with periodic boundary conditions.}

\label{fig:AB_disorder}

\end{figure}

In Fig.~\ref{fig:AB_disorder} shows the effect of disorder on the spectrum of the non-Hermitian Aharonov-Bohm cage from Fig.~\ref{fig:lattice}.  These results are obtained by numerically diagonalizing Eq.~\eqref{eq:rspace_disorder} for various disorder strengths $W$; the other lattice parameters are chosen so that the lattice is tuned to EP$_1$ in the disorder-free limit ($W = 0$).  As described in Section~\ref{sec:lattice}, the ordered lattice has flat bands at energies $E_1 = 0$ (a flat band of non-Hermitian degeneracies), and $E_2=\Delta=2C$ (a Hermitian-like flat band).  In Fig.~\ref{fig:AB_disorder}(a), we plot the mean disorder-induced shifts $|\delta E| \equiv |E - E_{1,2}|/C$ to the energy eigenvalues as a function of the disorder strength $W$. The shifts grow as $\delta E \sim \sqrt{W}$ in the non-Hermitian flat band, and $\delta E \sim W$ in the Hermitian-like flat band. In both cases, $\mathrm{Re}(\delta E)$ and $\mathrm{Im}(\delta E)$ have similar magnitudes.

To characterize the degree of localization of the eigenstates $\Psi_n^{(j)}$ in the disordered lattice, we compute the participation numbers $P_j = 1/\sum_n |\Psi_n^{(j)}|^4$.  The results are shown in Fig.~\ref{fig:AB_disorder}(b). For the compact localized states of the non-Hermitian flat band, $P_{\mathrm{CLS}}=2$ (in comparison, a delocalized state has $P \sim N$, where $N$ is the lattice size). For both the bands, we find $P_j \sim P_{\mathrm{CLS}}$ regardless of the value of $W$.  This indicates that the eigenstates of a weakly-disordered lattice behave much more like compact localized states than a delocalized Bloch wave basis.  The disorder simply causes the states to be exponentially localized, rather than vanishing identically outside the localization region.

This behavior is consistent with previous studies on the effect of diagonal disorder on isolated one-dimensional Hermitian flat bands~\cite{polariton_flat_band,leykam_FB}.  There, the vanishing of the band velocity ($v_G = 0$) implies that the disorder strength $W$ sets the only relevant energy scale. Applying degenerate perturbation theory, one can show that the energy bandwidth is set by $W$, while the eigenstates themselves are insensitive to $W$ (always being in the ``strong disorder'' regime $W \gg v_G$).  Our present results show that non-Hermitian flat bands exhibit similar behaviour.  Interestingly, however, the presence of non-Hermitian degeneracies results in a square root sensitivity of the energy eigenvalues to perturbations, $\delta E \sim \sqrt{W}$~\cite{EP,exceptional_point}. This holds as long as $W$ is smaller than the band gap; larger $W$ will induce hybridization between the bands, affecting the mode profiles. For example, in the strong disorder limit $W \gg C$ the inter-waveguide coupling will become negligible, resulting in $P_j \approx 1$. 

\begin{figure}

\includegraphics[width=0.55\columnwidth]{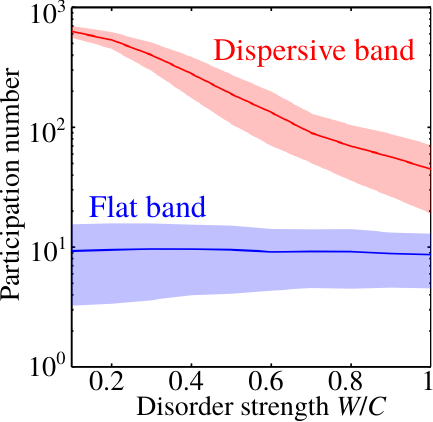}

\caption{Mean participation number versus disorder strength $W$ for the disordered embedded flat band lattice, showing the eigenstates at the flat band energy ($E=0$; blue) and in the center of the upper dispersive band ($E \approx 2.8C$).  The results are averaged over 20 disorder samples, and shaded regions denote the standard deviation.  Here $\gamma = 2C/3$ and $\Delta = 2C$, as in Fig.~\ref{fig:embedded_FB}, and the lattice size is $N=500$ unit cells with periodic boundary conditions.}

\label{fig:embedded_disorder}

\end{figure}

Fig.~\ref{fig:embedded_disorder} shows the effects of disorder on the ``embedded'' non-Hermitian flat band from Fig.~\ref{fig:embedded_FB}. In Hermitian systems, the compact localized states of embedded flat bands are known to be unstable against disorder: hybridization with dispersive band states with nonzero group velocity results in delocalization of the eigenmodes as $W \rightarrow 0$, i.e. $P \sim 1/W \rightarrow \infty$~\cite{leykam_FB}. Here, in contrast, the flat band states have $P \sim P_{\mathrm{CLS}}$ as $W \rightarrow 0$ (blue curve), while eigenstates in the middle of the dispersive band (red curve) exhibit conventional delocalization. This shows that localization in the embedded non-Hermitian flat band is also robust to weak disorder.

Finally, we note that off-diagonal disorder (i.e., random couplings induced via imperfections in the waveguide or resonator spacings) preserves the bipartite sublattice symmetry of our model. In this case, the flat band eigenstates will remain at $E=0$, protected by the sublattice symmetry.  The eigenstate profiles will be weakly perturbed, but remain compactly localized~\cite{mukherjee2015,polariton_flat_band,chiral_FB}.

\section{Conclusion}

We have shown how to generalize flat bands to non-Hermitian systems by combining a bipartite symmetry with frustration between energy conserving (Hermitian) and non-conserving (non-Hermitian) couplings. Such systems are ``frustrated'' because Hermitian couplings favour eigenstates with neighboring sites either in-phase or $\pi$ out-of-phase (to ensure zero net power flow and a stationary intensity distribution), whereas the non-Hermitian couplings favor $\pm \pi/2$ phase differences (to balance power flow between media with gain and loss).  At first glance, the energy spectra and compact localized states of the non-Hermitian flat bands resemble that of their Hermitian counterparts.  However, deeper analysis of the eigenstates and propagation dynamics reveals important differences. For example, the non-Hermitian flat bands can host isolated exceptional points which are unaccompanied by a $\mathcal{PT}$-breaking transition.  The compact localized states may be subexponentially amplified, and robust to weak disorder.

In the future, it would be interesting to further study, both theoretically and in experiments, the peculiar beam dynamics associated with the non-Hermitian flat band states, along with their interplay with disorder and interactions. In settings where Aharonov-Bohm caging induced by real or synthetic magnetic fields is difficult to implement, the non-Hermitian flat bands may provide a useful alternative method to design flat dispersion and achieve transverse confinement of light. 

\begin{acknowledgments}
This research was supported by the Singapore National Research Foundation (grant NRFF2012-02), the Singapore MOE Academic Research Fund Tier 2 (grant MOE2015-T2-2-008), and the Institute for Basic Science through Project Code (IBS-R024-D1).
\end{acknowledgments}


\begin{thebibliography}{99}

\bibitem{bergholtz_review}
E.~J. Bergholtz and Z. Liu, {\it Topological flat band models and fractional Chern insulators}, Int. J. Mod. Phys. B {\bf 27}, 1330017 (2013).

\bibitem{derzhko_review}
O. Derzhko, J. Richter, and M. Maksymenko, {\it Strongly correlated flat-band systems: The route from Heisenberg spins to Hubbard electrons}, Int. J. Mod. Phys. B {\bf 29}, 1530007 (2015).

\bibitem{frustration}
R. Moessner and A. P. Ramirez, {\it Geometrical frustration}, Phys. Today {\bf 59}, 24 (2006).

\bibitem{optical_lattice}
G.-B. Jo, J. Guzman, C.~K. Thomas, P. Hosur, A. Vishwanath, and D.~M. Stamper-Kum, {\it Ultracold atoms in a tunable optical kagome lattice}, Phys. Rev. Lett. {\bf 108}, 045305 (2012); J. Struck, C. \"Olschl\"ager, M. Weinberg, P. Hauke, J. Simonet, A. Eckardt, M. Lewenstein, K. Sengstock, and P. Windpassinger, {\it Tunable gauge potential for neutral and spinless particles in driven optical lattices}, Phys. Rev. Lett. {\bf 108}, 225304 (2012).

\bibitem{coupled_lasers}
M. Nixon, E. Ronen, A.~A. Friesem, and N. Davidson, {\it Observing geometric frustration with thousands of coupled lasers}, Phys. Rev. Lett. {\bf 110}, 184102 (2013).

\bibitem{ab_cage}
J. Vidal, R. Mosseri, and B. Dou\c{c}ot, {\it Aharonov-Bohm cages in two-dimensional structures}, Phys. Rev. Lett. {\bf 81}, 5888 (1998).

\bibitem{ab_cage_experiment}
C. C. Abilio, P. Butaud, Th. Fournier, B. Pannetier, J. Vidal, S. Tedesco, and B. Dalzotto, {\it Magnetic field induced localization in a two-dimensional superconducting wire network}, Phys. Rev. Lett. {\bf 83},  5102 (1999).

\bibitem{vicencio2015}
R. A. Vicencio, C. Cantillano, L. Morales-Inostroza, B. Real, C. Meijia-Cortes, S. Weimann, A. Szameit, and M. I. Molina, {\it Observation of localized states in Lieb photonic lattices}, Phys. Rev. Lett. {\bf 114}, 245503 (2015).
%

\bibitem{mukherjee2015}
S. Mukherjee, A. Spracklen, D. Choudhury, N. Goldman, P. \"Ohberg, E. Andersson, and R. R. Thomson, {\it Observation of a localized flat-band state in a photonic Lieb lattice}, Phys. Rev. Lett. {\bf 114}, 245504 (2015).
% 

\bibitem{diamond_ladder}
S. Mukherjee and R.~R. Thomson, {\it Observation of localized flat-band modes in a quasi-one-dimensional photonic rhombic lattice}, Opt. Lett. {\bf 40}, 5443 (2015).

\bibitem{polariton_flat_band}
F. Baboux, L. Ge, T. Jacqmin, M. Biondi, A. Lemaitre, L. Le Gratiet, I. Sagnes, S. Schmidt, H. E. T\"ureci, A. Amo, and J. Bloch, {\it Bosonic condensation and disorder-induced localization in a flat band}, Phys. Rev. Lett. {\bf 116}, 066402 (2016).

\bibitem{sawtooth}
S. Weimann, L. Morales-Inostroza, B. Real, C. Cantillano, A. Szameit, and R. A. Vicencio, {\it Transport in sawtooth photonic lattices}, Opt. Lett. {\bf 41}, 2414 (2016).

\bibitem{pt_lattice_1}
K. G. Makris, R. El-Ganainy, and D. N. Christodoulides, {\it Beam dynamics in PT symmetric optical lattices}, Phys. Rev. Lett. {\bf 100}, 103904 (2008).

\bibitem{pt_lattice_3} A.~Szameit, M.~C.~Rechtsman, O.~Bahat-Treidel, and M.~Segev, \textit{$PT$-symmetry in honeycomb photonic lattices}, Phys.~Rev.~A \textbf{84}, 021806(R) (2011).

\bibitem{pt_lattice_4} H.~Ramezani, T.~Kottos, V.~Kovanis, and D.~N.~Christodoulides, \textit{Exceptional-point dynamics in photonic honeycomb lattices with $PT$ symmetry}, Phys.~Rev.~A \textbf{85}, 013818 (2012).

\bibitem{pt_optical_lattice}
A. Regensburger, C. Bersch, M.-A. Miri, G. Onishchukov, D.~N. Christodoulides, and U. Peschel, {\it Parity-time synthetic photonic lattices}, Nature {\bf 488}, 167 (2012).

\bibitem{zhen2015}
B. Zhen, C.~W. Hsu, Y. Igarashi, L. Lu, I. Kaminer, A. Pick, S.-L. Chua, J.~D. Joannopoulos, and M. Solja\v{c}i\'{c}, {\it Spawning rings of exceptional points of out Dirac cones}, Nature {\bf 525}, 354 (2015).

\bibitem{cerjan2016}
A. Cerjan, A. Raman, and S. Fan, {\it Exceptional contours and band structure design in parity-time symmetric photonic crystals}, Phys. Rev. Lett. {\bf 116}, 203902 (2016).

\bibitem{bender}
C.~M. Bender and S. Boettcher, {\it Real spectra in non-Hermitian Hamiltonians having PT symmetry}, Phys. Rev. Lett. {\bf 80}, 5243 (1998).

\bibitem{PT_review}
K.~G. Makris, R. El-Ganainy, D.~N. Christodoulides, and Z.~H. Musslimani, {\it PT-symmetric periodic optical potentials}, Int. J. Theo. Phys. {\bf 50}, 1019 (2011).

\bibitem{PT_review_2}
V.~V. Konotop, J. Yang, and D.~A. Zezyulin, {\it Nonlinear waves in PT-symmetric systems}, Rev. Mod. Phys. {\bf 88}, 035002 (2016).

\bibitem{pt_breaking}
A. Guo, G. J. Salamo, D. Duchesne, R. Morandotti, M. Volatier-Ravat, V. Aimez, G.~A. Siviloglou, and D.~N. Christodoulides, {\it Observation of PT-symmetry breaking in complex optical potentials}, Phys. Rev. Lett. {\bf 103}, 093902 (2009).

\bibitem{pt_optics}
C.~E. R\"uter, K.~G. Makris, R. El-Ganainy, D.~N. Christodoulides, M. Segev, and D. Kip, {\it Observation of parity-time symmetry in optics}, Nature Phys. {\bf 6}, 192 (2010).

\bibitem{EP}
W.~D. Heiss and H.~L. Harney, {\it The chirality of exceptional points}, Eur. Phys. J. D {\bf 17}, 149 (2001).

\bibitem{exceptional_point}
T. Gao, E. Estrecho, K.~Y. Bliokh, T.~C.~H. Liew, M.~D. Fraser, S. Brodbeck, M. Kamp, C. Schneider, S. H\"ofling, Y. Yamamoto, F. Nori, Y.~S. Kivshar, A.~G. Truscott, R.~G. Dall, and E.~A. Ostrovskaya, {\it Observation of non-Hermitian degeneracies in a chaotic exciton-polariton billiard}, Nature {\bf 526}, 554 (2015).

\bibitem{ding_2015}
K. Ding, Z. Q. Zhang, and C. T. Chan, {\it Coalescence of exceptional points and phase diagrams for one-dimensional PT-symmetric photonic crystals}, Phys. Rev. B {\bf 92}, 235310 (2015).

\bibitem{EP_acoustic}
K. Ding, G. Ma, M. Xiao, Z.~Q. Zhang, and C.~T. Chan, {\it Emergence, coalescence, and topological properties of multiple exceptional points and their experimental realization}, Phys. Rev. X {\bf 6}, 021007 (2016).

\bibitem{invisibility}
S. Longhi, {\it Invisibility in non-Hermitian tight-binding lattices}, Phys. Rev. A {\bf 82}, 032111 (2010).

\bibitem{invisibility_2}
Z. Lin, H. Ramezani, T. Eichelkraut, T. Kottos, H. Cao, and D.~N. Christodoulides, {\it Unidirectional invisibility induced by PT-symmetric periodic structures}, Phys. Rev. Lett. {\bf 106}, 213901 (2011).

\bibitem{EP_laser}
B. Peng, S. K. \"Ozdemir, M. Liertzer, W. Chen, J. Kramer, H. Yilmax, J. Wiersig, S. Rotter, and L. Yang, {\it Chiral modes and directional lasing at exceptional points}, Proc. Natl. Acad. Sci. USA {\bf 113}, 6845 (2016).

\bibitem{EP_laser_2}
P. Miao, Z. Zhang, J. Sun, W. Walasik, S. Longhi, N.~M. Litchinitser, and L. Feng, {\it Orbital angular momentum microlaser}, Science {\bf 353}, 464 (2016).

\bibitem{lin2016}
Z. Lin, A. Pick, M. Loncar, and A. W. Rodriguez, {\it Enhanced Spontaneous Emission at Third-Order Dirac Exceptional Points in Inverse-Designed Photonic Crystals}, Phys. Rev. Lett. {\bf 117}, 107402 (2016).

\bibitem{chern2015}
G.-W. Chern and A. Saxena, {\it PT-symmetric phase in kagome-based photonic lattices}, Opt. Lett. {\bf 40}, 5806 (2014).

\bibitem{pt_flat}
L. Ge, {\it Parity time symmetry in a flat band system}, Phys. Rev. A {\bf 92}, 052103 (2015).

\bibitem{molina2015}
M.~I. Molina, {\it Flatbands and PT-symmetry in quasi-one-dimensional lattices}, Phys. Rev. A {\bf 92}, 063813 (2015).

\bibitem{PT_FB}
H. Ramezani, {\it Non-Hermiticity-induced flat band}, Phys. Rev. A {\bf 96}, 011802(R) (2017).

\bibitem{lieb}
E.~H. Lieb, {\it Two theorems on the Hubbard model}, Phys. Rev. Lett. {\bf 62}, 1201 (1989).

\bibitem{chiral_FB}
A. Ramachandran, A. Andreanov, and S. Flach, {\it Chiral flat bands: existence, engineering and stability}, arXiv:1706.02294 (2017).

\bibitem{leykam_FB}
D. Leykam, J.~D. Bodyfelt, A.~S. Desyatnikov, and S. Flach, {\it Localization of weakly disordered flat band states}, Eur. Phys. J. B {\bf 90}, 1 (2017).

\bibitem{ab_cage_longhi}
S. Longhi, {\it Aharonov-Bohm photonic cages in waveguide and coupled resonator lattices by synthetic magnetic fields}, Opt. Lett. {\bf 39}, 5892 (2014).

\bibitem{EP_continuum}
S. Longhi, G. Della Valle, {\it Optical lattices with exceptional points in the continuum}, Phys. Rev. A {\bf 89}, 052132 (2014).

\bibitem{active_coupler}
N. V. Alexeeva, I. V. Barashenkov, K. Rayanov, and S. Flach, {\it Actively coupled optical waveguides}, Phys. Rev. A {\bf 89}, 013848 (2014).

\bibitem{dark_state_laser}
C.~M. Gentry and M.~A. Popovic, {\it Dark state lasers}, Opt. Lett. {\bf 39}, 4136 (2014).

\bibitem{longhi_gauge}
S. Longhi, D. Gatti, and G. Della Valle, {\it Non-Hermitian transparency and one-way transport in low-dimensional lattices by an imaginary gauge field}, Phys. Rev. B {\bf 92}, 094204 (2015).

\bibitem{dark_laser_experiment}
H. Hodaei, A.~U. Hassan, W.~E. Hayenga, M.~A. Miri, D.~N. Christodoulides, and M. Khajavikhan, {\it Dark-state lasers: mode management using exceptional points} Opt. Lett. {\bf 41}, 3049 (2016).

\bibitem{longhi2016}
S. Longhi, {\it Non-Hermitian tight-binding network engineering}, Phys. Rev. A {\bf 93}, 022102 (2016).

\bibitem{diffusive}
S. Mukherjee, D. Mogilevtsev, G. Y. Slepyan, T. H. Doherty, R. R. Thomson, and N. Korolkova, {\it Coherent diffusive photonics}, arXiv:1703.06025 (2017).


\end{thebibliography}
\end{document}